%% file: paper.tex
\documentclass[12pt]{article}

\usepackage{setspace}
\usepackage{fullpage}
\usepackage{amsmath}
\usepackage{graphicx}%
\usepackage{amsfonts}%
\usepackage{amssymb}
\usepackage{epsfig}
\usepackage{times}
\usepackage{url}
\usepackage{rotating}
\usepackage{caption}
\usepackage{enumitem}
\setlist{nolistsep,leftmargin=*}

\usepackage{algorithmic}
\usepackage{algorithm}

\newcommand{\comment}[1]{} 

\title{{\bf Results from a Practical Deployment of the MyZone Decentralized P2P Social Network} \\
\it Tech. Report}
\author{ {\bf Alireza Mahdian, Richard Han, Qin Lv and Shivakant Mishra}  \\
Department of Computer Science \\
University of Colorado at Boulder\\
{\small alireza.mahdian@colorado.edu}
}
\date{}

\begin{document}

\pagestyle{plain}
\pagenumbering{roman}
\maketitle

% Use the following at camera-ready time to suppress page numbers.
% Comment it out when you first submit the paper for review.
%\thispagestyle{empty}
\pagebreak

\begin{abstract}
This paper presents MyZone, a private online social network for
relatively small, closely-knit communities. MyZone has three important
distinguishing features. First, users keep the ownership of their data
and have complete control over maintaining their privacy. Second, MyZone
is free from any possibility of content censorship and is highly
resilient to any single point of disconnection. Finally, MyZone minimizes 
deployment cost by minimizing its computation, storage and network bandwidth
requirements. It incorporates both a P2P architecture and a
centralized architecture in its design ensuring high availability, security
and privacy. A prototype of MyZone was deployed over a period of 40 days
with a membership of more than one hundred users. The paper 
provides a detailed evaluation of the results obtained from this deployment.
\end{abstract}

\pagebreak
\tableofcontents
\pagebreak

\cleardoublepage
\pagenumbering{arabic}

\input{intro.tex}

\input{related.tex}

\input{assumptions.tex}

\input{design.tex}

\input{evaluation.tex}

\input{contributions.tex}

\small
\bibliography{refs}
\bibliographystyle{plain}

\pagebreak

\end{document}

%% file: intro.tex
\section{Introduction}
\label{ch:intro}

%poularity/usefuleness of OSN
%Centralized architecture
	%adv: ease of management (policy ...); high avaiabiility
	%issues: privacy, censorship
%P2P
	%issue: security, availability, 
	%Structured vs unstructured
	%Problems with structured
%Myzone description
%Contributions

%Online social networks (OSNs)such as facebook, twitter and linkedIn have gained
%immense popualrity over the last decade. They offer enormous benefits
%in terms of facilitating users to expand their social circles through common
%interests, mutual friends or even searching for long lost acquaintances, to
%share content with different users either individually or as groups, to
%allow users to express opinions on shared contents, and much more. These
%networks typically employ a centralized architecture wherein central servers
%provides both storage and connectivity services.
 
Online social networks (OSNs) are typically designed as centralized systems,
where central servers provide both storage and connectivity services.
The key advantages of this design include high availability of user
profiles (irrespective of whether they are online or not), ease in 
establishing 
online interconnectivity among users, and ease of management in terms of
enforcing providers' security and privacy policies. Indeed, all popular OSNs
such as Facebook, Twitter and LinkedIn have been built as centralized systems.
However, as some of the recent events have shown
\cite{facebook_call, cnn_egypt_censor, epic_fail, washingtonpost, cnn_government, nytimes_china_censor}, users of these systems
lose control of their privacy and ownership of their data, and
the systems themselves are prone to censorship and complete disconnection.
Furthermore, building a centralized server incurs large resource and
management cost.

In this paper, we propose MyZone, a private online social network 
for facilitating social communication using standard social network
features among the
members of a relatively small, closely-knit community. MyZone provides three
distinguishing characteristics. First, users keep the ownership of their data
and have complete control over maintaining their privacy. In particular,
users are not resigned to store their profiles at a central server owned
by OSN providers, and neither are they required to store them in unknown
storage nodes typically determined by DHT-style techniques in P2P 
networks. Instead, users in MyZone store their data on their own devices
and replicate them on the devices of the users that they trust. MyZone
provides support for efficient propagation of profile updates to all replicas,
and maintains the privacy and integrity of user data despite being replicated
in multiple independent administrative domains.

Second, MyZone is free from any possibility of content censorship. Users
own their content, and no single entity or a small group of entities can
conspire to block them. Furthermore, MyZone is highly resilient to any
single point of disconnection. It remains operational even when it is
under active assault from a single entity or a small group of entites trying
to shut it down. Finally, MyZone employs a (mostly) decentralized 
architecture. While a centralized server is needed for initial user
registrations and (sometimes) to locate friends' profiles, the bulk of
communication takes place in a P2P fashion, and most importantly, 
the centralized server is not used for storing user profiles. Hence, computation,
storage and network bandwidth requirements of centralized servers are
drastically reduced. This significantly reduces the resource or maintenance
cost of deploying MyZone.

\begin{figure} \begin{center}\includegraphics[width=0.5\textwidth]{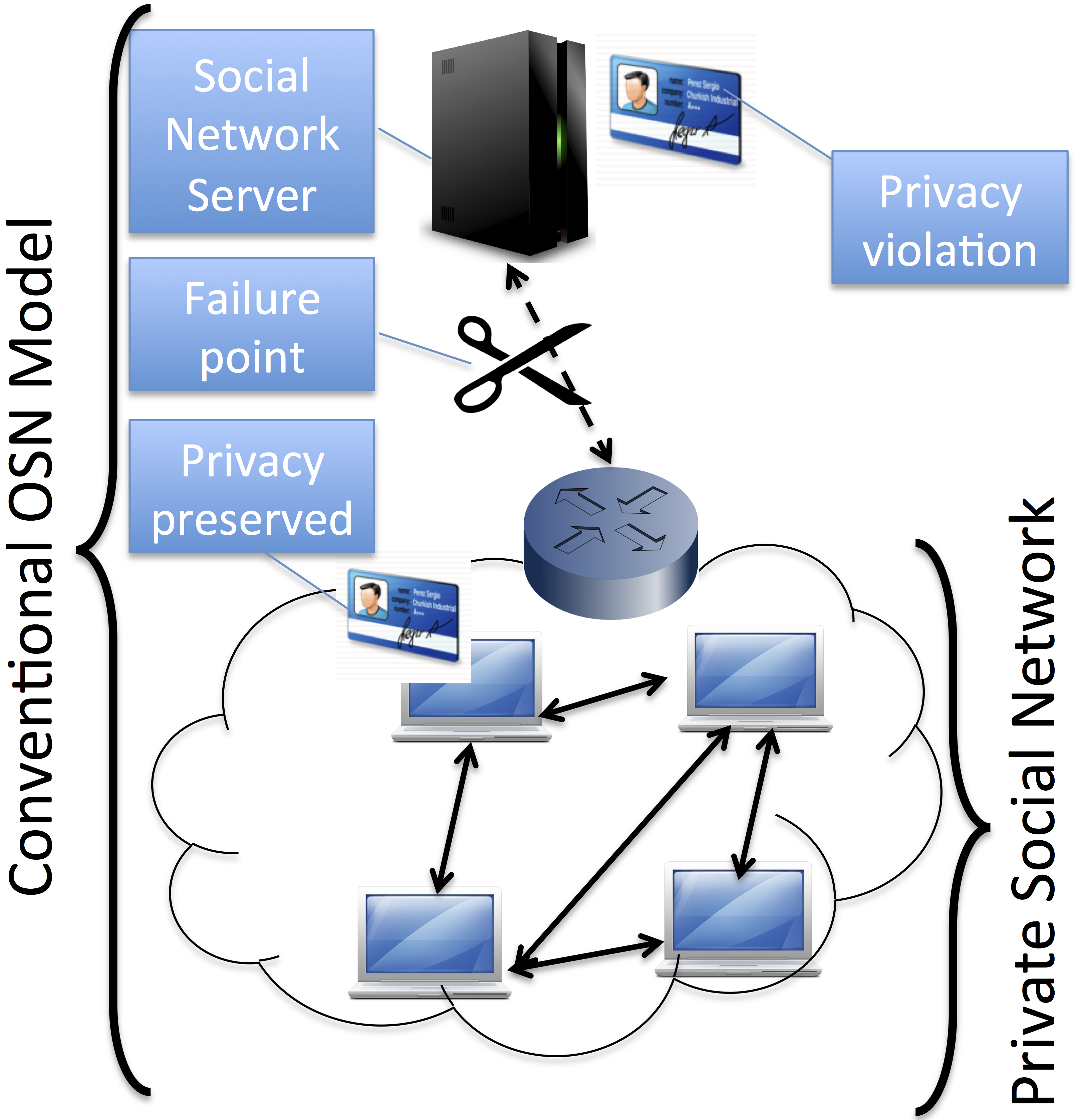} \end{center}
\caption{Depiction of conventional OSNs' deficiencies. \label{fig:motivation} }
\end{figure}

While MyZone can certainly be used for building a large-scale OSN, our 
focus in this paper is to facilitate private OSN for relatively-small
(hundreds to thousands of members) and closely-knit communities.  Friends
who do not wish to expose their data to  third-party social network
providers such as Facebook or Google+ can form their own private social networks
using MyZone.  Also, MyZone is very well suited for building communities
that wish to keep their data confidential, yet continue to organize, e.g., an OSN for
high school students, parents and teachers to facilitate social as well as academic
interactions among themselves.  \comment{This may include sharing ideas about a class assignment,
notifications from teacher's homework or gradebook page, planning for 
a senior picnic, etc.  

Another example is an OSN for a neighborhood community
involving people living within a few blocks. Members may use such a network 
for social acivities as well as community planning activities.
}  MyZone is appropriate for organizing activities such as democratic
advocacy in countries where the state may easily deny access to traditional
social networks by cutting off Internet access to the servers of those networks. Figure \ref{fig:motivation} illustrates the shortcomings of conventional OSNs and how our private social network fits in the overall picture.

%close friends as members
%no access to large resources
 
MyZone incorporates both a P2P architecture and a centralized
architecture in its design. It
employs an unstructured P2P network for storing user 
profiles. A user's profile is stored on his/her device and mirrored on
the devices of a few other users based on the user's trust. MyZone ensures
that user profile is only accessible to the user's friends and any updates
such as writing on the wall are quickly propagated to all mirrors. 
In addition, MyZone utilizes a central server for user registration and for
storing some important meta information. A prototype of MyZone has been
implemented. The paper reports the results of an extensive deployment 
of this prototype that included more than one hundred members over a period 
of 40 days.

The rest of this paper is organized as follows. Section \ref{ch:related}
reviews some important related works and Section \ref{ch:assumptions} states
the system assumptions and challenges in building MyZone. 
Section \ref{ch:systemDesign} describes the detailed system design and Section 
\ref{ch:evaluation} describes the prototype deployment and evaluates MyZone
based on the results of this deployment. Finally, Section
\ref{ch:contributions} concludes the paper.

%% file: related.tex
\section{Related Work}
\label{ch:related}

There have been several attempts at P2P social networks, motivated by the
desire to achieve user privacy. Safebook \cite{safebook} is a decentralized
and privacy-preserving OSN that provides two types of overlays: a set of
concentric structures called {\it matryoshkas} created around each node to
provide data storage and communication privacy; and a P2P substrate providing
lookup services. Matryoshkas are built using the idea of selecting trusted 
friends as mirrors. There are several issues with the design of Safebook.
First, a node can join Safebook only by invitation. Second,
the reachability of a user on Safebook depends on the number of mirrors.
Third, offline messages like wall posts are not accessible until the user
comes back online and publishes those updates. Fourth, Safebook is dependent
on a third-party P2P system to implement a DHT to find entry points for each
node. Another serious problem with Safebook is that it does not address
friend revocation that is an inherent part of any OSN. Finally, using the
overlay matryoshkas would result in longer paths on the IP infrastructure
yielding poor performance. 

A high level description of another P2P OSN called PeerSoN is provided
in \cite{peerson}. PeerSoN has a two-tier
architecture: a look up tier that uses DHT and a second tier that consists
of peers and contains user data. It is implicitly assumed that the users
are connected most of the time, perhaps through different devices and from
different locations. It is not clear if and how these devices would be
synced with each other. Offline messages are stored on a DHT, which means scalability may be an issue here. Furthermore, user profiles are not replicated,
which means a user's profile is not accessible if she is offline. Finally,
friend revocation is not considered as a functionality of the system. 

Vis-$\grave{a}$-Vis\cite{vis-a-vis} is an OSN design that uses Virtual Individual Servers (VIS) as personal virtual machines running in a paid compute facility. In Vis-$\grave{a}$-Vis, a person stores her data on her own VIS, which arbitrates access to that data by others. VISs self-organize into overlay networks corresponding to social groups. Vis-$\grave{a}$-Vis is designed so that it can interoperate with existing OSNs. The main issue with this system is that users have to pay for a VIS and in many cases take charge of maintenance of their own VIS.

Cuckoo\cite{cuckoo} is a microblogging decentralized social network that
uses a centralized OSN only as backup and as a byproduct saving bandwidth
costs and reducing downtime while performing equally well. Cuckoo organizes
user clients into a structured P2P overlay using Pastry.
Cuckoo does not provide any encryption and can only be applied to
microblogging services like Twitter with very limited functionalities.

In addition, a variety of past and ongoing projects on P2P social networking
include Diaspora\cite{diaspora}, The Appleseed Project\cite{appleseed},
and Peerbook\cite{peerbook} that all require an online server. Obtaining a
publicly accessible server is not free of charge, and needs regular monitoring and maintenance on behalf of the user.
All P2P OSNs to date assume public IP addresses for peers and they do not
address the problem of NAT traversal. Furthermore, with a exception of a few, the majority of these works have not been implemented while none have been thoroughly evaluated in real world deployments. 

%In addition, there is existing work on secure and anonymous P2P networking, e.g. Darknet\cite{darknet}, GNUnet\cite{gnunet}, I2P\cite{iip}, Bunzilla\cite{bunzilla}, and Freenet\cite{freenet}.  In spite of all these projects, and in spite of all the obstacles facing conventional OSNs, still it is safe to say that P2P social networking is far from being considered as a threat to OSN providers Due to their distributed nature, in large measure because the aforementioned P2P OSNs lack major competing features compared to centralized OSNs. 

%At the current state, the trend seems to indicate that users would prefer features and convenient user experience at the cost of privacy violations. Therefore, we believe that P2P OSNs would only be able to compete with their centralized counterparts if they can provide the same features and functionalities at the same level of user experience.  MyZone is designed to ultimately achieve the goal of providing competitive features offered by conventional OSNs while supporting key benefits like privacy, security, and resiliency.

%% file: assumptions.tex
\section{System Model}
\label{ch:assumptions}

\comment{
In this section we first state the system properties that MyZone requires in order to function. This is followed by the definition of the security model composed of {\bf 1)} a trust model that describes the trusted entities in the system and their trust relations with each other, and {\bf 2)} an adversary model that identifies different adversaries, their resources, and the kind of attacks that they can implement.

\subsection{System Requirements}
\label{sec:systemRequirements}

We divide the system requirements into two categories: participating devices and network infrastructure. }

We begin by making the following assumptions about the system and network connectivity.  Participating devices can be desktops, laptops or smartphones with network connectivity either through Ethernet or wireless.  Although each device can be behind any kind of firewall or NAT, a reasonably small number of devices should be able to obtain public IP addresses.  Out of all those devices with public IP addresses, at least one, should have two network connections, i.e., dual homed, with different public IP addresses.  Furthermore, a couple of those participating devices with public IP addresses, should be able to host databases that serve other peers.  We assume there exists a functioning IP infrastructure at all times, even though it may be partitioned to some extent either intentionally or due to unpredictable outages.  

Given these assumptions, our design goals are the following:  (1) MyZone should support all primary functionalities of a conventional OSN like Facebook, including the ability to establish and revoke friendships, sharing contents and comments with friends at different levels, and participating in discussions; (2) It should preserve user privacy by ensuring that users own their own data, and no central authority or third party can access user information without explicit permission from the user;  (3) It should be highly resilient in face of malicious attacks as well as infrastructure failures caused by unpredictable causes, possessing self healing properties by automatically reconciling the states after connection has been restored;  (4) It should tolerate a high rate of churn due to mobile users, and seek to provide near 24/7 availability of user profiles (best effort); and (5) \comment{Due to the growing trend of switching to smartphones and tablet computers, the underlying computing platform consists of traditional desktops and laptops with high bandwidth network connections, as well as portable devices with restricted resources. To deal with the latter, }The system should be designed to be minimally demanding of the limited resources of portable devices, and should incorporate power/resource management techniques.

We choose a decentralized P2P model based on user trust to realize the above design goals of our system, rather than a traditional client-server model.  Users will own their own social data and store it on their own machines, including their own desktops, laptops, and mobile devices, or on the devices of others they trust, rather than storing their social data on an untrusted third-party server or on a DHT.  Thus, friends may act as mirrors for each other, thereby allowing updates to a user's profile even if the user's primary machine is not currently accessible, as may be the case for mobile devices.   In order to protect communications and authenticate users, MyZone employs cryptographic methods as well as a certificate authority so that only users within a zone may participate in that zone.  We show through the description of our design as well as the evaluation of a real world deployment of our prototype application, that such a trust-based decentralized system can be designed to support standard social network functions, preserve user privacy, be highly resilient to benign failures and malicious attacks, tolerate high mobile churn, and achieve these objectives while conserving resources.

%Make the case for decentralized social networks here�

%[our P2P model actually devolves into a server model if there are no mirrors�]

The challenges that we face in achieving the design goals of MyZone span six key areas, namely trust, availability, resiliency, routing, connectivity, and resource efficiency.  In this paper, we will focus primarily on the areas of trust and availability, but mention the other challenges later in this section.  

%Then talk about the challenges due to decentralization: main ones are trust, availability, resiliency, etc�

%Put back much of the content in old Sections 3 and 4 here...

\subsection{Trust Model}
\label{subsubsec:trustmodel}

The trust-based challenges introduced by our private social network concern identifying and maintaining the different levels of trust encountered within the system.  A user will form a relationship of trust with one or more friends.  In this case, the user $U$ can write on the social network profiles of each of his/her friends, and each of his/her friends can write on user $U$'s profile.  Within user $U$'s friends, certain friends will be trusted even more to act as mirrors in case user $U$'s own devices are unavailable to receive profile updates, e.g.,  posting to $U$'s wall.  In this case, the mirrors will absorb these update requests, and then re-synchronize with user $U$ when user $U$ is reconnected to the network.  In fact, we need to distinguish between five levels of trust in MyZone:
\begin{itemize}
\item {\it Trust as a friend}: User $A$ trusts user $B$ as her friend and gives $B$ read and write access to her profile. This trust is symmetric but not transitive.
\item {\it Trust as a mirror}: User $A$ trusts user $B$ as her mirror and grants $B$ permission to serve as $A$'s mirror whenever $A$ is unavailable. User $A$ already trusts $B$ as a friend, hence, this is a stronger level of trust.
\item {\it Trust as a replica}: In this scenario, user $A$ is a friend of user $C$ and user $B$ is a mirror for user $C$. User $A$ trusts user $B$ to act as a benevolent mirror on behalf of user $C$ whenever $C$ is unavailable. 

This kind of trust can be referred to as ``{\it trust by proxy}'' and it is in a way, derived based on a transitive relationship on ``{\it trust as a friend}'' and ``{\it trust as a mirror}''. One important property of this type of trust, is one directional read and write access, i.e.,  only $A$ can read from and write to $C$'s replicated profile on $B$, and $B$ cannot modify or read $A$'s profile. 

\item {\it Trust as a certificate authority}: Users put their utmost trust in $CA$ to act as a certificate authority for the private social network. The $CA$ issues certificates to all participating users, and users can verify all the certificates that are issued by the $CA$. 

We assume that the $CA$ is not penetrable by malicious nodes, while it can be the target of other attacks. Also, the $CA$ will never act maliciously during its lifetime. These assumptions are essential for the OSN to function\comment{, and everything falls apart if one of these assumptions fails}.  
\item {\it Trust as a user}: User $A$ can verify the identity of user $B$. This can be done through a certificate issued to $B$ by a certificate authority (CA) that is trusted by $A$.
\
\end{itemize} 

We need to preserve these levels of trust in the face of a variety of security threats.  We consider three types of adversaries that may be present in the system.  First, some users may be  {\it ``honest but curious''} who want to view the user profiles of non-friend users.  Second, there may be malicious entities that are not users of the private social network, but want to attack the system in any of the following ways: eavesdropping; spoofing; (distributed) denial of service; IP or URL filtering; and backtracking, which is an attempt to link some content to a user, e.g., a government agency trying to find the user who is responsible for posting some particular content.  Third, there may be insider threats from malicious users who are part of the private social network.  In general, malicious entities have two motivations, namely to gather information about users and use it against individuals or the social network as a whole, e.g., implementing a divide and rule policy among users by spreading mistrust, and to prevent access to the social network, and hence, create an outage to a subset of the system.

We assume that two friends will not act as adversaries of each other. This does not apply to friends of friends, since trust is not transitive. We also assume that a benevolent mirror is trusted with the integrity of data (it will not maliciously modify the data) but it may act to gain profile information from its clients, i.e., mirrors can be ``honest but curious'' users.

We will therefore seek a design that prevents the disclosure of profile information to any user who is not a friend.  A user profile can only be viewed and modified by her friends.  MyZone is designed to be resilient against DDoS attacks and URL/IP filtering.  Also, every modification of a user profile is authenticated.

\comment{
\subsection{Challenges}
\label{sec:challenges}

The shortcomings of existing OSNs, demand a distributed architecture for the next generation of OSNs. Based on the distributed architecture, the assumptions mentioned earlier, and the design goals, we identify six key challenges in the design, implementation, and deployment of
MyZone: availability, resiliency, routing, connectivity, security, and traffic optimization and power management.
}

\subsection{Availability Challenges}
\label{subsec:availability}

An important property of OSNs is that a user profile is always available regardless of whether that user is currently online or not.  In MyZone, we seek to preserve this property by employing mirroring of a user's profile amongst a user's trusted friends.  We assume that a user selects friends based on social trust via out-of-band discussions.  Under these conditions, our challenges include the following:

\begin{enumerate}
\item When and how profile updates are propagated to other mirrors.
\item How to maintain replica consistency among all mirrors.
\item What type of replica consistency is appropriate for this application.
\item Whether a user profile is completely or partially replicated at different mirrors.
\item Whether the mirrors themselves should be prioritized as primary, secondary,  and so on. 
\item How many mirrors are needed for a desired level of availability?
\end{enumerate}

MyZone is designed so that user profiles are ``almost always'' available.  A weakly consistent view of the profile is always available.  Given the amount of mobile churn, we felt that the consistency and availability guarantees should be best effort, as the benefits gained through absolute availability and total consistency are not worth the added complexity in the design of the OSN.

\subsection{Other Challenges}

A variety of other challenges remain, including networking and resource challenges.  We face some practical engineering challenges in terms of the network, such as the fact that many users may have their profiles hosted on peers that are positioned behind different NATs, i.e., full cone, address-restricted cone, or peer restricted cone.  To bridge these connectivity challenges, we introduce a relay server into the architecture, as described in the next section.  In terms of resource challenges, many users' primary devices may be mobile, which are resource-constrained.   A challenge is to develop a replication strategy that takes into account these limitations while providing availability and resiliency. One strategy is to leverage more resource-rich devices, i.e., online desktops, to support the bulk of mirroring duties, and to use mobile mirroring only as a strategy of last resort.  Heterogeneity of device resources and their limitations call for mirroring strategies that take into account the limited and diverse memory, bandwidth
and energy of the various peers comprising the private social network.

%% file: design.tex
\section{System Design Architecture}
\label{ch:systemDesign}

%This section gives a detailed description of our design. 
MyZone's design is based on a two layered architecture.  The lower layer referred to as the {\it service layer} provides essential services to the components of the system and facilitates the registration of peers, finding peers, establishing connections between peers and much more. 
%These services are designed to provide a reliable, resilient and secure infrastructure addressing the requirements of P2P applications, from NAT traversal, to reliable UDP connections, to secure socket services, and more. Hence, the service layer can be used by any P2P application that requires these features and is not specific to just OSNs. 
The upper layer is referred to as the {\it application layer}. This layer provides standard social network features such as wall postings, etc.  In addition, it is responsible for implementing higher level security policies such as read and write permissions for particular users or groups of users. Profile replication is also done at this layer. 
%Next we describe the two layered architecture starting from the lower layer.

\begin{figure} \begin{center}\includegraphics[width=0.8\textwidth]{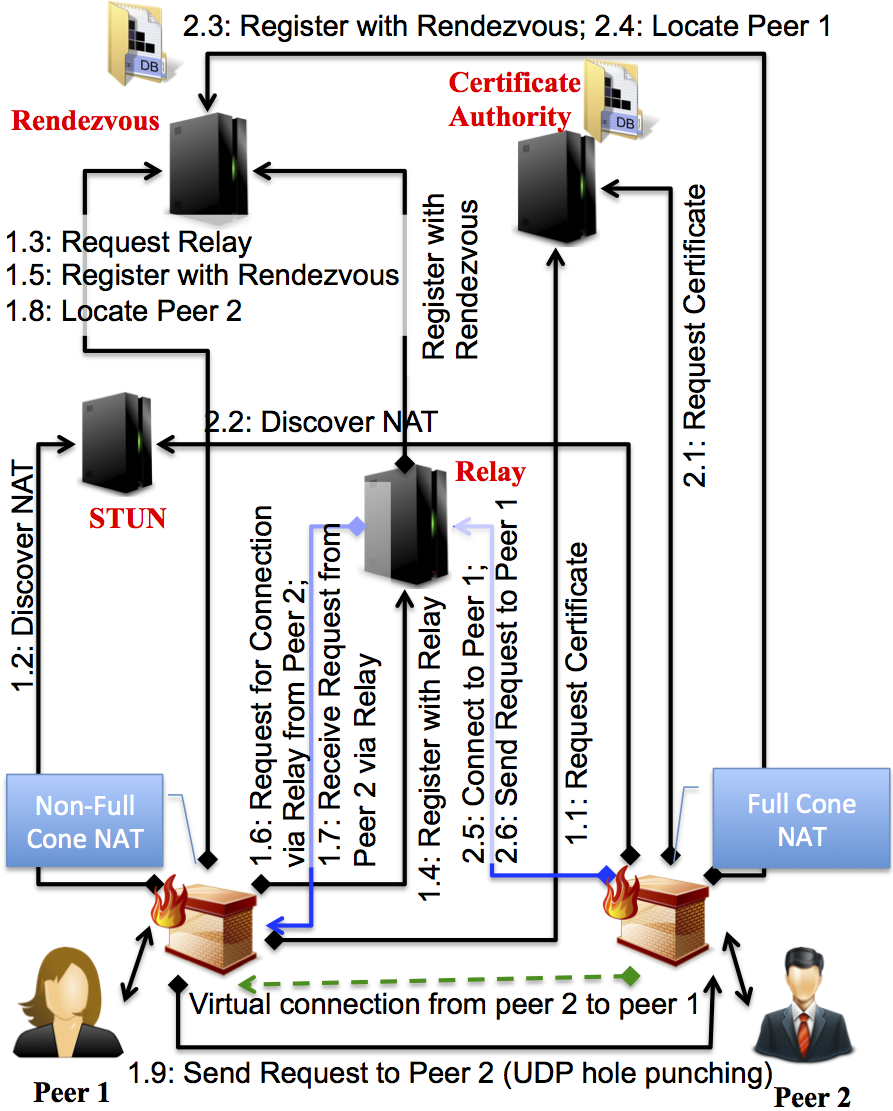} \end{center}
\caption{Deployment diagram for a private social network. \label{fig:local_deployment} }
\end{figure}

\subsection{Service Layer}
\label{sec:serviceLayer}

Figure \ref{fig:local_deployment} illustrates the major steps in terms of how MyZone establishes connections between peers using the rendezvous server and a certificate authority.  The sequence of steps for each peer is represented as $i.x$ where $i$ is the peer and $x$ is the step.  Each new user needs to first obtain a certificate from the CA server, which verifies that the usernames are unique.  We assume that peers have already obtained the public key of the CA which is needed to verify the certificates. 
Then, when peers come online, they need to securely register with rendezvous servers in order to be discovered by other peers.   When a peer comes online and wishes to find a friend, they securely query the rendezvous server to find the address of the friend in order to connect with another peer.  Recall that the rendezvous server is completely trusted in the model of a private social network, so it is appropriate for it to know the addresses of the peers within its zone.   \comment{WHY IS THIS NOT AN INVASION OF PRIVACY???}
%The server should be able to host a database.
%This database stores the information for peers as well as relays. 
%In case, a peer needs a relay to accept connections, it will query the rendezvous server for a relay server.  
%The CA server would issue certificates to each new user and verifies that the usernames are unique. We assume that peers have already obtained the public key of the CA which is needed to verify the certificates. Note that several CAs can coexist in the system and a user can have several certificates issued by different CAs. 
Two other components, namely a STUN server and a set of relay servers are also needed to deal with NAT connectivity issues. Our architecture handles attacks on the rendezvous server by periodically synchronizing the rendezvous server database with sever other backup rendezvous servers that are all concurrently online and available. Any DDoS attack on a rendezvous server is handled by changing the IP address of the server under attack and spreading the IP address to all users through the backup rendezvous servers. The rendezvous servers essentially collaborate with each other to create a dynamic decentralized server.

The service layer provides an infrastructure that is resilient against partitioning, reliable even on top of UDP, and secure against malicious attacks. It is also very dynamic and can be used by peers behind all kinds of firewalls and NATs. In this section we describe the service layer by describing its components and their interactions with each other.  For the sake of brevity, we focus our discussion on the two major steps of  the protocol, namely registering a peer and locating another peer.  Details of other steps of the protocol, such as obtaining a valid CA-signed certificate, registration of the relay servers with the rendezvous server, the involvement of the STUN server, etc. can be found in our technical report~\cite{MyZoneTechReportARXIV}.

Peers need to be able to register with a rendezvous server in order that other peers can find them and
establish a P2P connection.
%Let's examine the steps that a peer should take, in order to establish a secure connection with another peer. A peer can directly connect to another peer if, 1) the IP address of the other peer is known, and 2) the peer either has a public IP address, or is behind a NAT that can be traversed using UDP hole punching. If the NAT traversal is not possible then the peer can only receive connections through a relay. 
%In order to be able to establish a secure connection between peers, they should be able to verify the identities of each other. This can be done by using certificates, which means, a certificate authority that is trusted by all peers needs to exist. 
%Then comes the important step of a peer, registering with the rendezvous server. 
%Figure \ref{fig:peer_registration_sd} represents the sequence of messages exchanged in this process 
The first message is sent by the peer and includes the peer's certificate, while the reply carries the session key generated by the rendezvous server, and encrypted by the public key of the peer. If the session key is correctly retrieved by the peer, it sends the following encrypted information to the rendezvous server: the priority of the device being registered on behalf of the user since a user may use several of her own devices as a method of self replication on the system; port number; type of NAT, e.g. full cone, non-full cone, or public IP address; type of protocol (TCP or UDP); IP address; relay server address and port if needed; a list of passphrases generated by the peer for each friendship that are unrelated to actual usernames but are used in lookup queries for each peer (known only to the peer's friends); list of mirrors of a peer including the usernames of its mirrors, which are implemented at the application layer as shown later; a signed message digest over the following information: IP address, port number, type of protocol, relay server address, relay server port and the list of passphrases to ensure the authenticity and the integrity of the data sent back by the rendezvous server; and a signed message digest over the list of mirrors.

The rendezvous server will store all this information in its database. Upon successful registration with the rendezvous server, the server returns a {\tt peer\_registered} reply along with an optional list of pending friendship requests. The friendship requests for peer $i$ are stored in form of friendship requester's username, and a passphrase specifically generated for $i$ encrypted using the public key of peer $i$. Encrypting the friendship requester's passphrase for $i$ is mandated based on a need to know basis, and prevents malicious entities, from looking up the IP address of the friendship requester, by providing the passphrase. 

\comment{
\begin{figure*}
\begin{center} \includegraphics[width=\textwidth]{locate_peer_sd.png} \end{center}
\caption{Sequence diagram of the scenario where a peer locates another peer. \label{fig:locate_peer_sd} }
\end{figure*}
}

%\subsubsection{Finding another peer}

Peer $i$ can look up the connection information for peer $j$, if 1) peer $j$ has already registered with the rendezvous server and, 2) peer $i$ has a passphrase with peer $j$.  A lookup query will first establish
a secure connection with the rendezvous server, retrieving a session key.  After the session key is retrieved, the peer sends the passphrase of the target peer to the rendezvous server. Upon finding a username corresponding to that passphrase in its {\it passphrase} table, the rendezvous server replies with the connection information of the target peer. The information includes the signed message digest described earlier, which will be used by the peer to verify the integrity and authenticity of the reply.

Once the IP address information of the other peer is retrieved then the peers can exchange signed
encrypted messages with one another, traversing NATs as needed~\cite{MyZoneTechReportARXIV}.

\comment{
\begin{figure} \begin{center} \includegraphics[width=\textwidth]{connecting_peers_sd.png} \end{center}
\caption{Sequence diagram of a scenario where a peer connects to another peer. \label{fig:connecting_peers_sd} }
\end{figure}

Finally, peer $1$ can connect to peer $2$ after obtaining its connection information using the sequence of messages described in Figure \ref{fig:connecting_peers_sd}. This process starts by peer $1$ sending its certificate to peer $2$. Peer $2$ would then send back its certificate encrypted using peer $1$'s public key. This encryption prevents an entity monitoring peer $1$'s traffic from figuring out the identity of peer $2$.

After the successful exchange of certificates, peer $1$ generates a session key and encrypts it first using peer $2$'s public key and then, its own private key. This double encryption ensures that the session key can only be retrieved by peer $2$ preventing any kind of attack compromising the session key. This single message has a crucial role in the security of the entire service layer. Peer $1$ can securely communicate with peer $2$ after the session key has been successfully retrieved by peer $2$.
}
  
\comment{
\begin{figure} \begin{center} \includegraphics[width=\textwidth]{send_friendship_request_sd.png} \end{center}
\caption{Sequence diagram of the scenario where a peer sends friendship request to a rendezvous server. \label{fig:send_friendship_request_sd} }
\end{figure}
}

Additionally, the service layer provides a method for a peer to send friendship requests for other peers. All friendship requests for a particular peer $x$ are sent to the rendezvous server, instead of sending them directly to $x$. There are two reasons for this. First, peer $x$ might not be online at the time of sending the friendship request and therefore may not receive it. Second, all friendship requests do not necessarily end in friendships and the connection information of peer $x$ should not be revealed to other peers unless $x$ has accepted their request and has a shared passphrase with them.  Due to space limitation we refer the reader to \cite{MyZoneTechReportARXIV} for a complete description of sending a friendship request.

\subsection{Application Layer}
\label{sec:applicationLayer}

\begin{figure} \begin{center} \includegraphics[width=\textwidth]{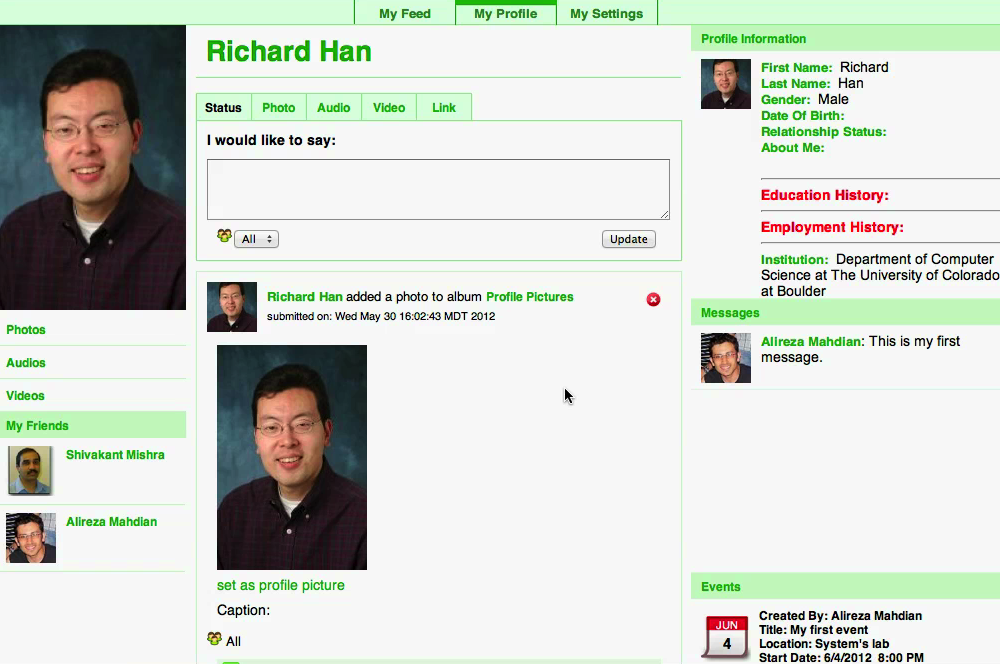} \end{center}
\vspace{-10pt}
\caption{A snapshot of MyZone's client application. \label{fig:MyZone} }
\vspace{-15pt}
\end{figure}

While the service layer is designed to overcome resiliency, connectivity, routing and most of security challenges\cite{MyZoneTechReportARXIV}, the application layer needs to tackle availability, traffic optimization and power management  challenges.  The application layer of MyZone is implemented on top of the service layer and provides the following functionalities:
\begin{enumerate}
\item Implementing all the social networking features of MyZone: Users can post links, status updates, video and audio files, and photos on their own or their friends' profiles. Any post can be deleted depending on the permission policy in place for that post.  Users can comment on, like or dislike any post.  Users can create events and invite their friends. Users can send private messages to their friends.  Users can assign their friends to customized {\it Zones} and share items with them exclusively.  The MyZone application UI is shown in Figure~\ref{fig:MyZone}.
\item Enforcing permission policies defined by users, for different elements of their profiles: These policies define read, write, or no access permissions, for each element of the profile. The user may define these access permissions, per individual, or groups of users. This is very similar to the concept of access control deployed in operating systems. 
\item Profile replication on other devices to increase availability.
\item Finally, if a user agrees to serve as mirror for a friend, the application layer needs to support the first two functionalities for the mirrored profiles.
\end{enumerate}

 \comment{We explain how the application layer would address these challenges as we describe the design. Before that we state all the features of MyZone that are going to be supported by the application layer. Most of these features are common on many social networking sites like Facebook and as a new competition are essential for the success of MyZone. MyZone has the following features:

\begin{itemize}
\item 
\item 
\item \item 
\item 
\end{itemize}
}

The MyZone user client is written as a cross-platform Java Web application, so it can be widely distributed on many platforms.  Social network profile information and updates are stored in platform-independent XML format.  In MyZone, every piece of information is mapped to either an attribute or an element.  This enables MyZone to support several client applications with various features and interfaces tailored to user specific requirements on the same social network. Each queryable piece of information has an id attribute that is unique and corresponds to the timestamp of the creation of that element. To achieve fast queries, each element is stored in ascending order in its corresponding XML file based on its id. Furthermore, to bound the time of a query each XML file only stores a limited number of elements determined by an attribute as part of the client application's settings. When querying for a particular element first the correct XML file is loaded into memory and then searched for the element with the correct id. Finding the correct file is done using a history log that keeps track of three items for each XML file: The id of the very first element stored in that file, the id of the very last element stored in that file and finally the number of items stored in the file. The history log itself is stored as an XML file as well.

To conserve traffic, the application layer uses a pulling scheme to distribute updates among users. This means that if two users $A$ and $B$ are friends with each other, $A$'s client application periodically tries to establish a connection to a device that is hosting $B$'s profile and upon successful connection, download the latest profile updates of $B$ onto $A$'s device. Upon every successful update session on $B$'s profile, $A$'s client application records the timestamp as the last update time for $B$. In order to optimize traffic, each update session for $B$'s profile initiated by $A$'s client application will be preceded by $A$'s client application sending out the last successful update time for $B$'s profile to $B$'s hosting device. The hosting device would then compute $B$'s latest profile updates for $A$ by combining all the updated elements of $B$'s profile since the last successful update time at $A$'s client application in a temporary XML file and sending it to $A$'s client application. This method minimizes traffic as each piece of profile update is transmitted as part of one and only one successful update session per friendship.

Avoiding retransmission of profile contents for each friend comes at the cost of storing friends' profiles locally. This means that at some point the client application may run out of storage for friends' profiles. As a measure to avoid this, the client application would store each of friends' profiles under a directory named after each friend's username in another directory appropriately named as {\it friends}. The {\it friends} directory has a restricted size, set as part of the client application's settings referred to as {\it cache size}. The application layer would fill the friends directory with friends' profile contents until the cache size capacity is reached. At that point, the latest profile updates will replace the oldest profile updates stored in the {\it friends} directory. Any deleted content can be retransmitted upon request. Therefore, there is a tradeoff between the storage and the traffic: as the cache size shrinks the amount of traffic due to retransmission of deleted profile content would increase.

Users can assign their friends to customized {\it Zones} e.g. friends, family, colleagues etc., and share items with them exclusively. These zones can be overlapping which means a friend can be assigned to several zones. When a friend is added, the client application automatically adds the new friend to a default zone called {\it All}. For every content a user posts on her profile, the corresponding XML element has a {\it sharedWith} attribute that implicitly dictates the read and write permissions of other friends for that content. Unless the content is shared with {\it All} zone, it will be only visible to members of the zone it is shared with. Furthermore, each posted content can be liked, disliked, or commented on by members of the zone that it is shared with. This feature provides more privacy to MyZone's users at the cost of more complexity especially when it comes to sending the latest profile updates to a friend $X$, in which case of all the latest profile updates, those that are only visible to $X$ should be sent back. This task is the most computationally intensive part of the client application.

The application layer increases profile availability when the client application is not connected by employing replication.  \comment{There are different design choices for replication. One solution is to replicate profiles on other devices either randomly or by using a heuristic approach. This approach is dominantly being used in many P2P systems such as bit torrent protocol and has proved to be very effective. But as we argued in \ref{subsec:availability}, this approach is not appropriate for MyZone due to the inherent nature of social networks and their features. Therefore we have invented our own replication scheme for MyZone.}   MyZone employs a trust-based replication scheme that will replicate user profiles on two classes of devices: devices that belong to the profile owner and devices that belong to friends of the profile owner. The replication on the first set of devices is referred to as {\it self replication} while replication on the second set of devices is labeled as {\it social hosting}. {\it Social hosting} is the concept of a friend accepting to be a mirror for another friend based on her social ties with the original profile owner.  
\comment{Based on the fact that social networks are formed on social ties, we believe that the process of mirroring in the social network should also be based on social ties. Mirroring for another user translates into allocating resources i.e. storage, cpu and bandwidth, on behalf of the user. It also translates into a sense of trust by the user in the mirror as the mirrored profile content includes sensitive information only available to friends. This sense of trust in addition to resource allocation required by the mirror would make the mirroring not just a role but a responsibility for the mirroring user.  This sense of responsibility can only be created among friends and not just any friend but those with stronger social ties to the original profile owner. Therefore we believe that social hosting is a more appropriate and effective method of replication for P2P social network applications such as MyZone. }

Social hosting is not forced onto friends. This means that a friend may or may not accept to be a mirror for another friend. This may cause a user not being able to obtain a mirror amongst her friends. In addition, a friend may limit the amount of storage he is willing to allocate as a mirror referred to as {\it mirroring capacity} as part of the mirror setting at any time. In this case a user may use self replication to increase the availability of her profile without any constraints until she can find an appropriate mirror. Self replication is done by assigning a priority to each device the user owns.  If a user owns a desktop, laptop and tablet/smartphone, then the user can use all three devices to host her profile, assigning the highest priority or rank to the desktop, while making the laptop secondary and the tablet her tertiary device. All requests to the user's profile are handled by the highest priority connected device.   \comment{Let's assume that the user $X$ has three devices at her disposal: A desktop at her home, a laptop that she uses at work and a tablet or smartphone that she carries with her all the time. Furthermore, all of these devices have some connectivity to the internet, while the desktop has most resources e.g. storage, power and cpu, the tablet has the least and the laptop is in between. User $X$ can use all three devices to host her profile by assigning the highest priority to the desktop making it her primary device, while making laptop her secondary and the tablet her tertiary devices. The device priority is otherwise known as the rank of the device. All the requests for $X$'s profile are handled by her desktop device first, and in case it is not online the laptop will handle the requests and otherwise the tablet. Note that the device priority is based on its resources and not its connectivity. When a higher priority device becomes online it will treat the lower priority devices as mirrors and will periodically synchronize with them.}

The rank of the device is generalized to social hosting as well. When a friend $Y$ of user $X$ agrees to mirror for $X$, $X$ will give a rank to $Y$. This rank can be a mixed function of $X$'s social tie with $Y$, the resources that $Y$ is allocating to mirror for $X$, and $Y$'s connectivity. While $X$'s devices are offline the requests for $X$'s profile are sent to her mirrors starting at the highest rank. This means that if a higher ranked mirror is available the lower ranked mirror is not going to be used. 

After $Y$ has agreed to be $X$'s mirror the most recent profile contents of $X$ will be sent to $Y$'s device as part of initial synchronization. In case $X$'s profile size is larger than the allocated capacity by $Y$, the mirroring space on $Y$ is filled with the latest contents from $X$'s profile and the old contents are not replicated. Upon successful initial synchronization, $X$ would send out $Y$'s username and its mirroring rank as part of its registration information to the rendezvous server\comment{ as illustrated in Figure \ref{fig:peer_registration_sd}}. The rendezvous server will store this information in its {\it mirrors} table\comment{described in Figure \ref{fig:rendezvous_db}}. This information will be distributed to all of $X$'s friend the next time that any of them tries to locate $X$\comment{ as illustrated in Figure \ref{fig:locate_peer_sd}}.

When a friend of user $X$ labeled as $Z$ tries to connect to $X$'s profile it will send a locate peer query to the rendezvous server with priority value of $0$ to connect to $X$'s primary device. Upon receiving the connection information, $Z$ tries to connect to $X$'s primary device. In case $X$'s primary device is unreachable $Z$ sends out a locate query for $X$'s secondary device by incrementing the priority value by one and repeats this process for the tertiary device in case the secondary device is also offline. Note that in our design we have limited the self replication to only two devices which means that a user can only have three devices associated with her MyZone profile and while a user must have a primary device, having secondary and tertiary devices is optional. When all three of $X$'s devices are unreachable, $Z$ can get the most recent profile updates of $X$ from her mirrors starting at the highest ranked mirror. To locate user $X$'s rank $i$ mirror, the value of priority sent as part of locate peer query \comment{in step 1.1.1.1 of Figure \ref{fig:locate_peer_sd}} is computed as $i+2$ where the offset by $2$ is indicating that the device is a mirror. 

\comment{
\begin{figure*} \begin{center} \includegraphics[width=\textwidth]{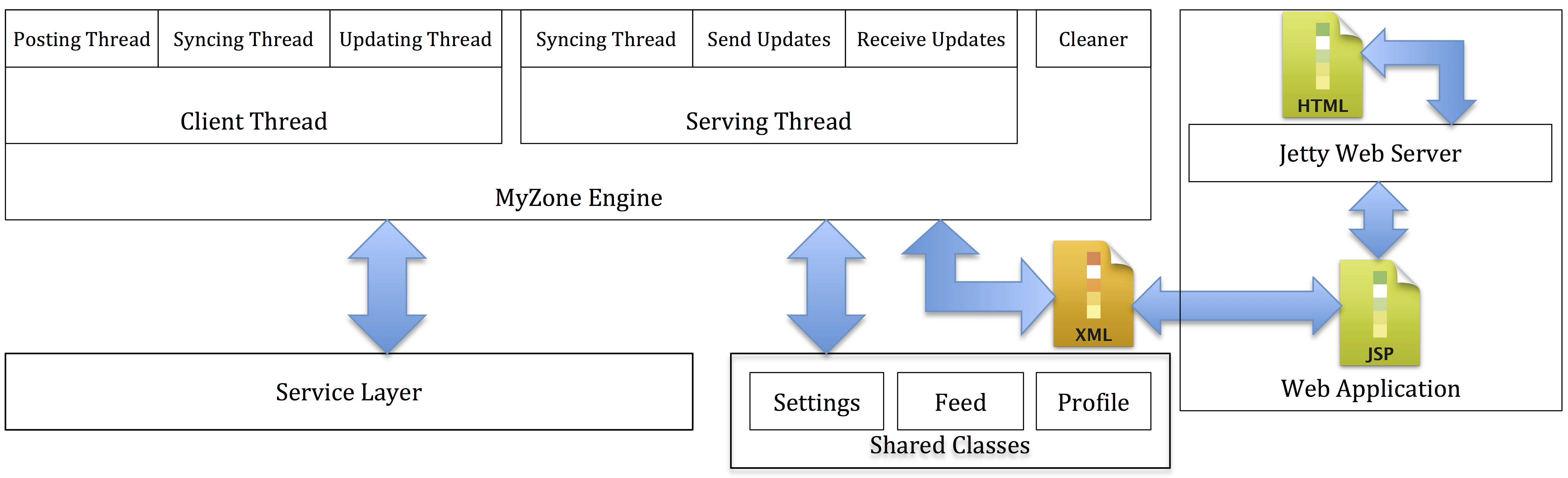} \end{center}
\caption{Design architecture for the entire application layer. \label{fig:design_architecture} }
\end{figure*}
}

%Figure \ref{fig:design_architecture} summarizes the three major components of the application layer: 
The application layer is realized as three major components: MyZone Engine, Web Application, and shared classes. The MyZone Engine is the most crucial part of the application layer. It runs as a background process and is responsible for downloading all the latest profile updates from friends, sending out all the contents posted by the user on friends' profiles, synchronizing with other replicas, and finally managing the storage capacity based on the limits set in the application setting. It is essentially responsible for all the P2P transmissions as well as storage management.   It consists of two threads.  A {\it client thread} periodically pulls the latest profile updates from other friends, sends out postings on other friends' profiles, and periodically synchronizes the user's latest profile updates with her mirrors. This is done by first receiving all the latest updates from the mirrors starting at the highest ranked mirror, merging those updates with the current profile and finally, sending out the updated profile to the mirrors.  A {\it server thread} receives requests for profile updates from a friend, and generates the correct delta update to that friend.  This involves bundling all the latest profile changes based on their {\it shareWith} attributes and the requester's membership in different zones.  The serving thread is also responsible for receiving and storing the updates posted by other friends on the user's profile.  Upon receiving a synchronization request from a mirrored profile owner, the serving thread is responsible for sending back all the latest updates of the mirrored profile and receiving the latest version of the mirrored profile from the original user.  The serving thread handles received requests for mirrored profiles as well as storage management of both friends' directory and all the mirrored profiles.

The Web application component of the application layer is responsible for representing all the information stored on XML files to the user through Web interfaces. The Web application layer includes a java based web server to provide the web interfaces. To convert XML files into Web pages we use JavaServer Pages (JSP) technology. The JSP code uses the classes in the shared class component of the application layer to convert XML files into Web pages. The user can view, modify or create new content using the Web interface. Finally, the shared classes component is  composed of a set of classes used as a toolbox by the other two components to convert the data between XML files and Java classes.

%% file: evaluation.tex
\section{Evaluation}
\label{ch:evaluation}

\comment{
Earlier in section \ref{sec:challenges}, we identified the challenges that a P2P social network faces. Furthermore, we explained how our design addresses those challenges in sections \ref{ch:systemDesign} and \ref{ch:securityMeasures}. While we theoretically showed how our design handles resiliency, routing, connectivity and security challenges, we did not produce any evidence of its effectiveness addressing availability and traffic optimization challenges. 

To demonstrate the feasibility of our design and evaluate its performance, we have implemented a fully-featured application. Using both real-world deployment and emulation, our evaluation focuses on three key aspects: {\em profile availability}, {\em resource utilization}, and {\em scalability}. 
} 

\comment{
Unless our design shows superior performance in all three aspects, it will not be able to successfully compete with existing centralized social networks. That is why In this section we thoroughly evaluate these aspects of MyZone using both real world deployment and emulation. We first describe our experimental setup, then we present the results of our experiments based on these aspects.
}

We have implemented a fully-featured application that was evaluated using both real-world deployment and emulation. Our evaluation focuses on three key aspects: {\em profile availability}, {\em resource utilization}, and {\em scalability}.

We ran a local deployment of MyZone for 40 days for a closely-knit
community of 104 users. Being in a closely-knit community, almost
all the users were friends with one another. We provided a rendezvous server, and a STUN server for the experiment in addition to 20 relay servers each capable of handling 20 concurrent connections. Client applications were configured to register with the rendezvous server every $2$ minutes and send their logs to the rendezvous server at the time of registration.  For each interaction initiated by a client the application created a single line in the log file that included the following information: {\it Session start time}, {\it Action type} (posting
and updates), 
{\it Target friend} (target friend's user name), {\it Serving friend}
(user name of the user hosting that friend's mirror), {\it Data size},
{\it Session status} (Connection established/not established,
successfully/unsuccessfully received updates, and successfully/unsuccessfully
sent posts), and {\it Session end time}.

\comment{
\begin{enumerate}
\item {\it Session start time}: Timestamp  (in millisecond) when a session started.
\item {\it Action type}: Posting on another friend's profile or getting updates from them.
\item {\it Target friend}: Hash value of the target friend's username, to anonymize the log data. 
\item {\it Serving friend}: Hash value of the username serving for the target friend. This hash value is the same as {\em target friend} if the connection is established to the device that belongs to the profile owner and is different if the connection is established to a mirroring friend of the original profile owner.
\item {\it Data size}: Number of bytes sent and received during this session.
\item {\it Session status}: Connection established, connection not established, successfully/unsuccessfully received updates and successfully/unsuccessfully sent posts.
\item {\it Session end time}: Timestamp (in millisecond) when a session ended.
\end{enumerate}
}

Table \ref{table:classifications} shows the geographical and timezone
distribution of the 104 users.
Before the start of our experiment, we provided a video tutorial of
MyZone to all users and also gave them a 5-day grace period to set up their basic profiles and add friends. The social graph was  almost complete with $5117$ edges and $104$ nodes. In order to evaluate MyZone under realistic loads we asked all users to duplicate their Facebook interactions on MyZone for the duration of our experiment. We also asked all users to create different zones if they wanted to and set the value of refresh interval for each zone based on their experience on Facebook. The average refresh interval reported by the users was $30$ minutes which meant that most of the users expected a new entry in their news feed approximately every $30$ minutes. 

$93\%$ of the users were behind symmetric firewalls which meant that they
had to use our relay servers in order to receive connections while the remaining $7\%$ used routers that supported UDP hole punching. We recorded $200$GB of traffic relayed over our servers over the duration of our experiment. Also, in order to measure the resiliency of MyZone against DDoS attacks on the rendezvous server we brought down the rendezvous server at the end of the 35th day of our experiment. Therefore, in the last 5 days of our experiment the client applications did not have access to the rendezvous server to locate their friends. 

\begin{table}
\caption{Geographic and Timezone Distribution of the Users.} \label{table:classifications}
\centering
\begin{footnotesize}
\begin{tabular}{|p{1.5cm}|p{1.5cm}|p{2cm}|}
\hline

Percentage of Total Subjects & Timezone & Continent\\
\hline
$59\%$& UTC -7:00& North America\\ 
\hline
$13\%$& UTC +3:30& Asia\\ 
\hline
$9\%$& UTC -4:00& North America\\ 
\hline
$9\%$& UTC -5:00& North America\\ 
\hline
$7\%$& UTC & Europe\\ 
\hline
$3\%$& UTC -6:00& North America\\ 
\hline
\end{tabular}
\end{footnotesize}
\end{table}

%\begin{figure}	
%	\centering
%	\begin{subfigure}[t]{2in}
%		\centering
%		\includegraphics[width=1.6in, height=1.4in]{s3f2.png}
%		\caption{Timezone classification of users}\label{fig:timezones}		
%	\end{subfigure}
%	\quad
%	\begin{subfigure}[t]{2in}
%		\centering
%		\includegraphics[width=1.6in, height=1.4in]{s3f3.png}
%		\caption{Geographical classification of users}\label{fig:geographic}
%	\end{subfigure}
%	\caption{Classification of users based on Timezone and Location}\label{fig:classifications}
%\end{figure}

\subsection{Availability}
\label{sec:availabilty}

Availability provides a measurement of how often a user profile is available. A
user profile is available when the machine that is hosting the profile is connected to the network. We compute average daily availability over the first 35 days, i.e., when the rendezvous server was available. 
More specifically, we consider update sessions (i.e., receiving profile updates from friends) and posting sessions (i.e., posting contents on friends' profiles). 
We define {\it success ratio} as the ratio of the number of successful sessions to the total number of sessions. Since each failed session results in a sequence of retries, we count a failed session and all of its consecutive retries only once to avoid overestimation of failed sessions.

\begin{figure} \begin{center} \includegraphics[width=\textwidth]{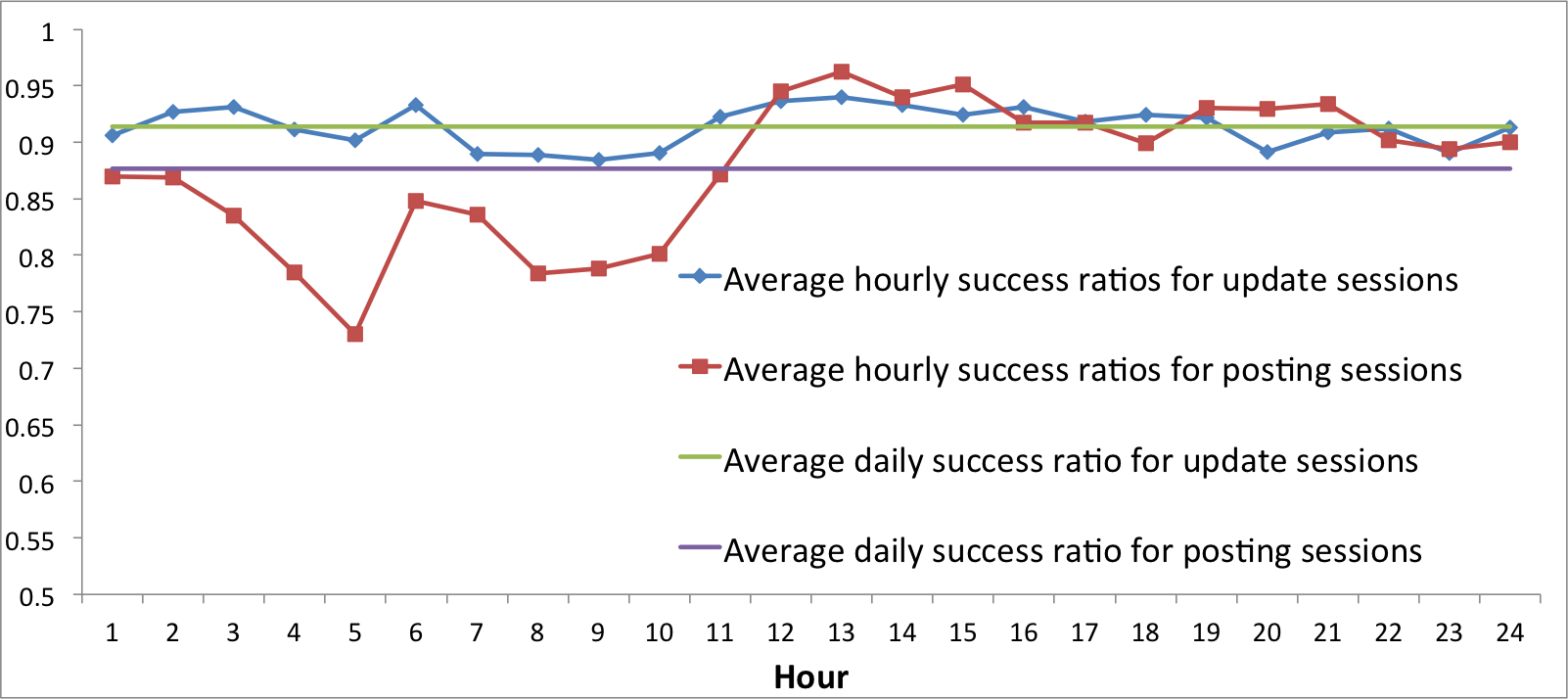} \end{center}
\vspace{-10pt}
\caption{Trend of average success ratios of update and posting sessions for a day. \label{fig:success_ratio_day} }
\vspace{-5pt}
\end{figure}

\begin{figure} \begin{center} \includegraphics[width=\textwidth]{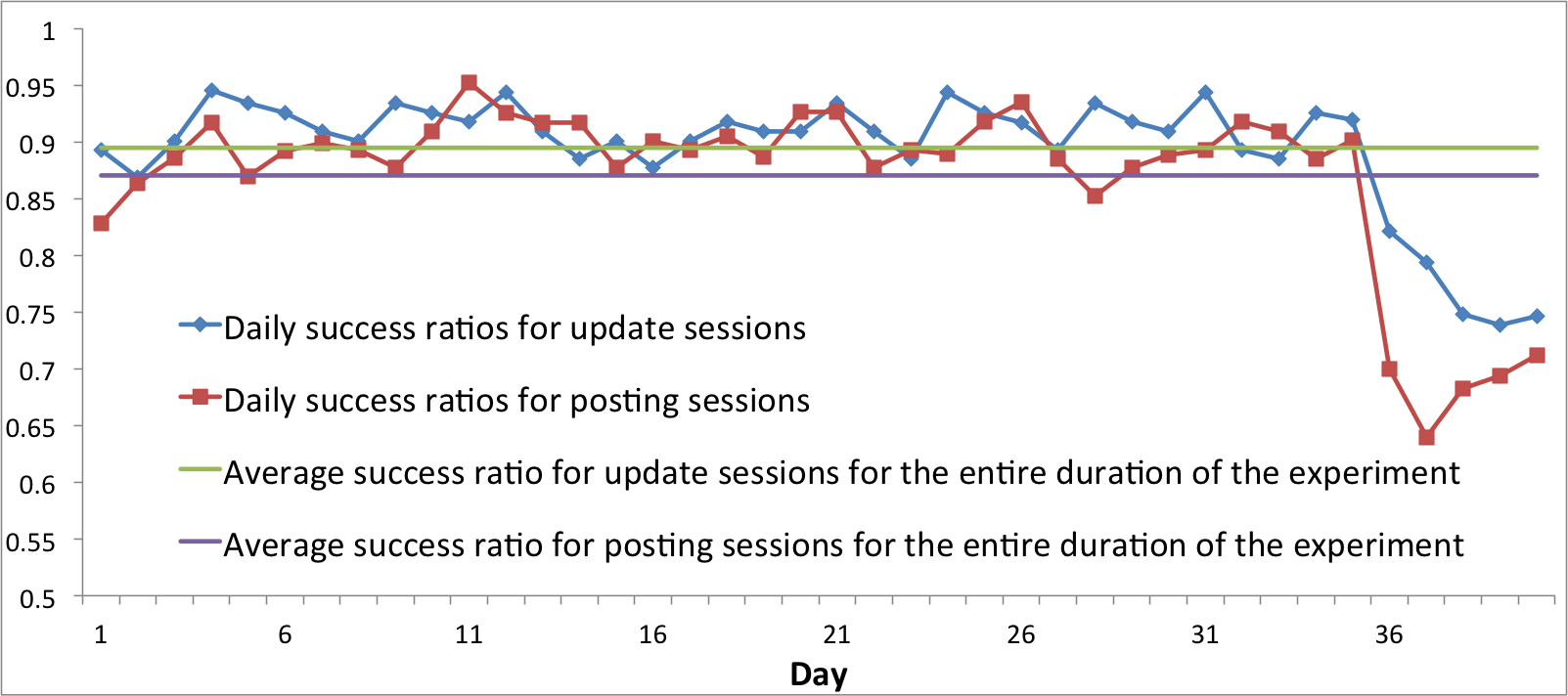} \end{center}
\vspace{-10pt}
\caption{Trend of success ratios of update and posting sessions for the entire duration of the experiment. 
\label{fig:success_ratio_all} }
\vspace{-15pt}
\end{figure}

Figure \ref{fig:success_ratio_day} shows the average success ratios of update and posting sessions for a day. The success ratios for both posting and update sessions are close to $90\%$ ($89\%$ for posting sessions and $92\%$ for update sessions). Success ratios were lower (still above 75\%) in the early hours of the day, when most users were offline. 
The success ratios for update sessions show smaller variations than those for posting sessions, since the number of update sessions is much larger than the number of posting sessions and the impact of failed sessions is relatively smaller. 
Figure \ref{fig:success_ratio_all} shows the daily success ratios for the entire duration of our experiment, which were consistently high for the first 35 days. Even for the last 5 days, when the rendezvous server was brought down, the drop was not very steep: down to 75\% for update sessions and 65\% for posting sessions. This shows that MyZone continues to be operational even when its
centralized server (rendezvous server) is brought down by adversaries. Also,
the average success ratios for the entire duration of our experiment are $89\%$ for update sessions and $87\%$ for posting sessions. 
\comment{ 
Although a drop in the success ratio is expected when the rendezvous server goes down, the slope of this drop is not that steep. This is contributed to the replication methods employed in MyZone. To put it in context, when a user device comes back online with a new connection it would be unreachable by her friends since they don't have the new connection information now that the rendezvous server is down, but, as the user obtains more mirrors either through friends or her own devices, the probability that one of her mirrors or other replicating devices has remained online hence having the same connection information would increase. Therefore the more mirrors a user has the more profile availability she gets even when the rendezvous server is down. 
}

\begin{figure}\vspace{-10pt} \begin{center} \includegraphics[width=0.6\textwidth]{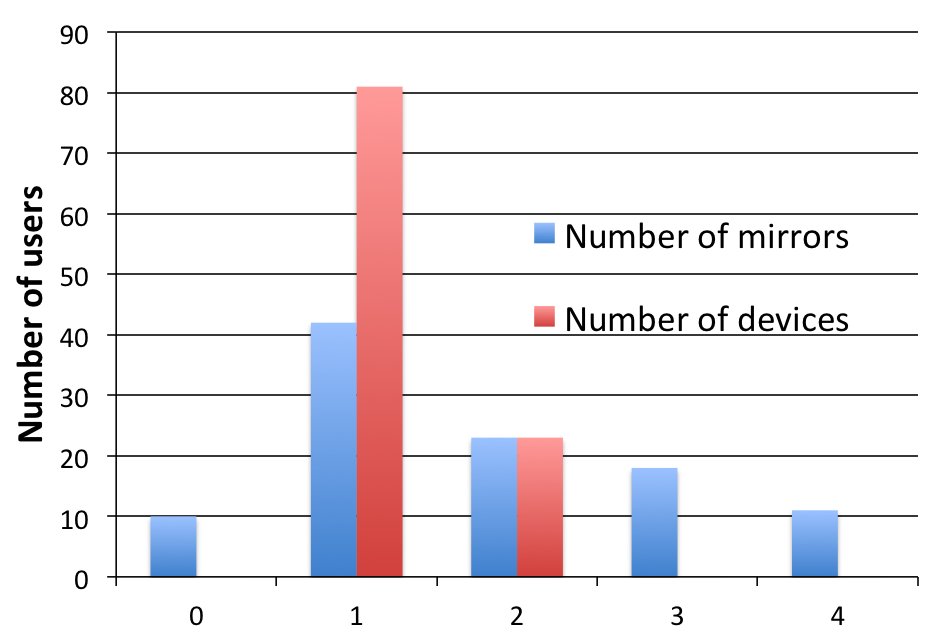} \end{center}
\vspace{-10pt}
\caption{Number of users based on their number of mirrors and devices. \label{fig:user_vs_mirrors_devices} }
\vspace{-5pt}
\end{figure}

Availability naturally depends on the number of devices and mirrors each user
has. Figure \ref{fig:user_vs_mirrors_devices} shows the distribution of the number of devices and mirrors for the users. Approximately $90\%$ of the users were able to obtain at least one mirror, while around $20\%$ had a secondary device available in addition to a primary device. Our results indicate that it was fairly easy for most of the users to obtain one mirror while obtaining more mirrors becomes harder. Note that mirroring in our design is purely based on the willingness of other friends and the strength of social ties between the users impacts this willingness a lot. That is why we refer to this method of mirroring as {\it social hosting}.

\begin{figure} \begin{center} \includegraphics[width=\textwidth]{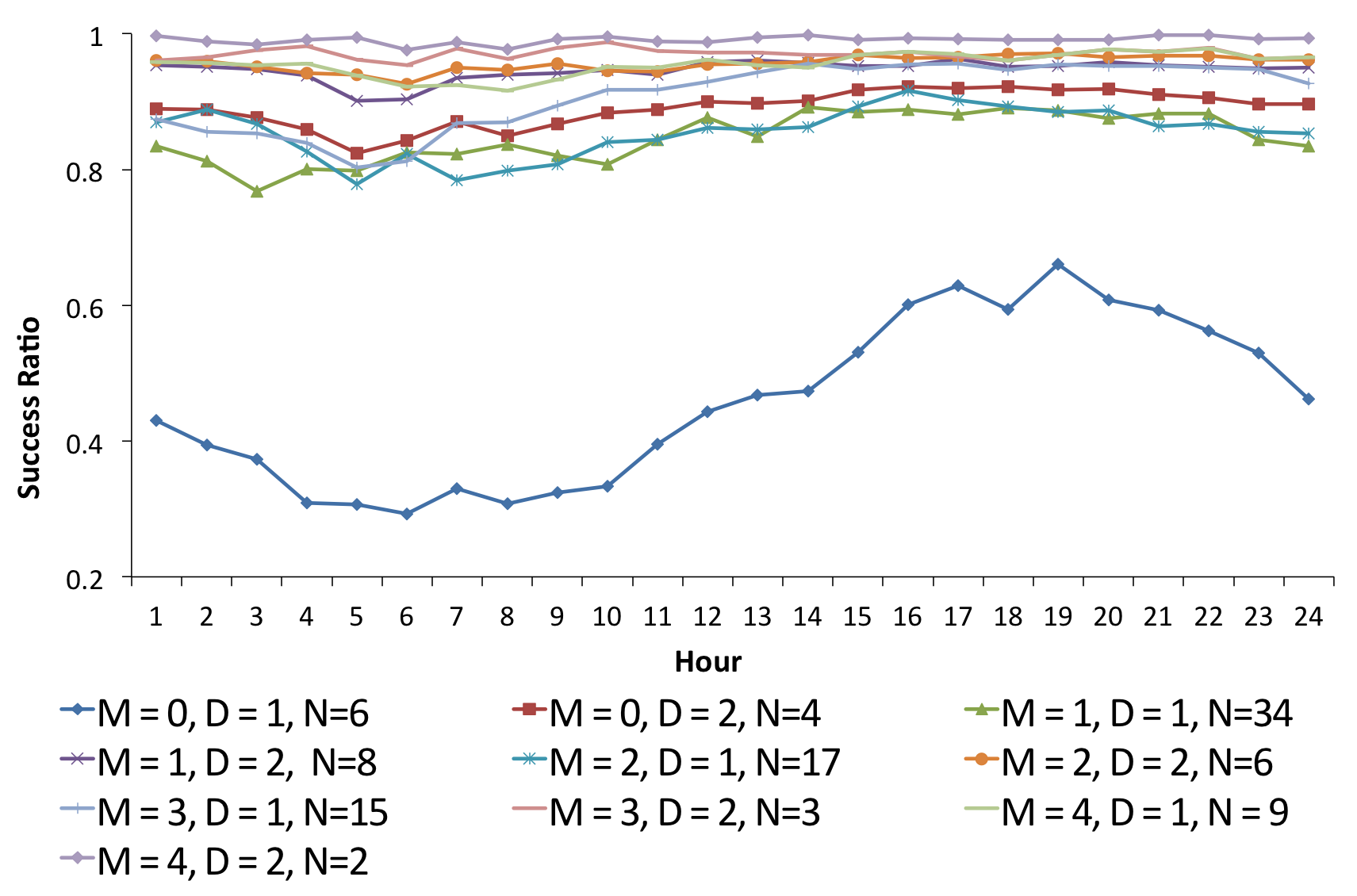} \end{center}
\vspace{-10pt}
\caption{Average success ratios of sessions over a day for different groups of users based on their number of mirrors and devices. M, D, and N denote the number of mirrors, devices and users respectively.\label{fig:user_vs_mirrors_devices_day} }
\vspace{-15pt}
\end{figure}

We categorize the users into different groups based on the number of their
devices and mirrors. 
Figure \ref{fig:user_vs_mirrors_devices_day} shows the number of users in each group as well as the average success ratios (of the first 35 days) for all sessions targeted for users of each group over a $24$ hour period starting at midnight. As expected, users with more mirrors and devices have higher profile availability. But the additional gain in the availability does not increase linearly with the number of mirrors and devices. Based on our results $1$ mirror and $2$ devices seems to result in optimal availability. 
\comment{ 
We can also see that the availability of the first group (i.e., no mirrors or additional devices, no replicas) is much less than that of the other groups. In fact out of the $6$ users in this group $2$ of them had nearly full time connectivity and that actually boosted the overall success ratio and if we exclude those two users from the group the average success ratio falls below $20\%$. 
} 

\begin{figure} \begin{center}\includegraphics[width=\textwidth]{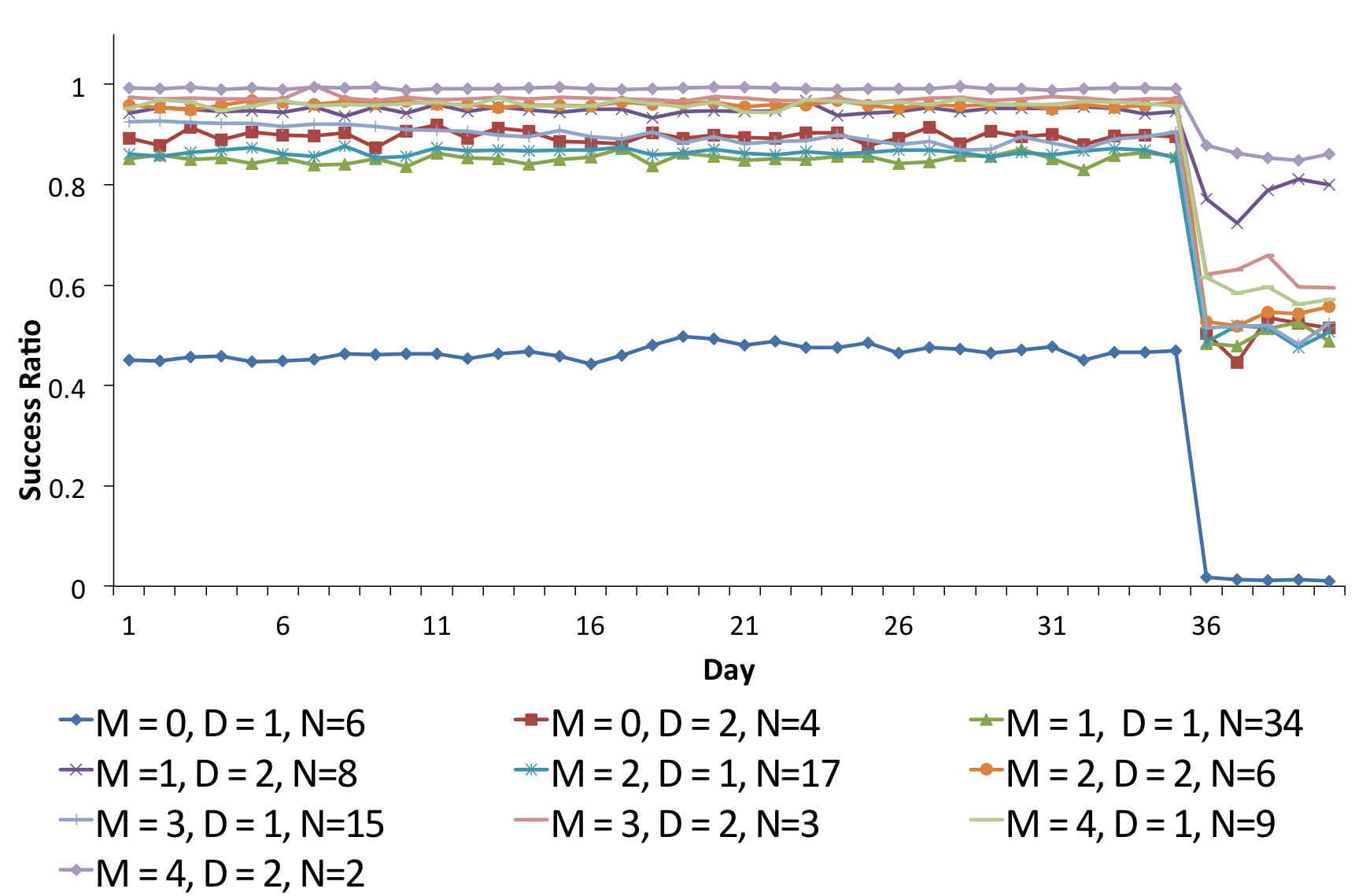} \end{center}
\vspace{-10pt}
\caption{Average success ratios of sessions over the entire duration of the experiment for different groups of users based on their number of mirrors and devices. M, D, and N denote the number of mirrors, devices and users respectively.\label{fig:user_vs_mirrors_devices_all} }
\vspace{-15pt}
\end{figure}

Figure \ref{fig:user_vs_mirrors_devices_all} shows the daily success ratios of all posting and update sessions for each group of users. We see that most of the groups maintained high availability. During the last 5 days, when the rendezvous server was down, the success ratios of users with $1$ mirror and $2$ devices is second only to those users with $4$ mirrors and $2$ devices. A careful analysis of the log files revealed that $3$ of the users belonging to this group have the same mirror which did not lose connectivity during the last $5$ days and hence it had the same connection information and therefore, even though the rendezvous server was out, user profile for those three users were always available through that mirror and hence the high overall success ratios for that entire group. Another observation is that the success ratio of the users with no replicas would drop to almost $0\%$ when the rendezvous server was down. 

\begin{figure} \vspace{5pt} \begin{center} \includegraphics[width=0.6\textwidth]{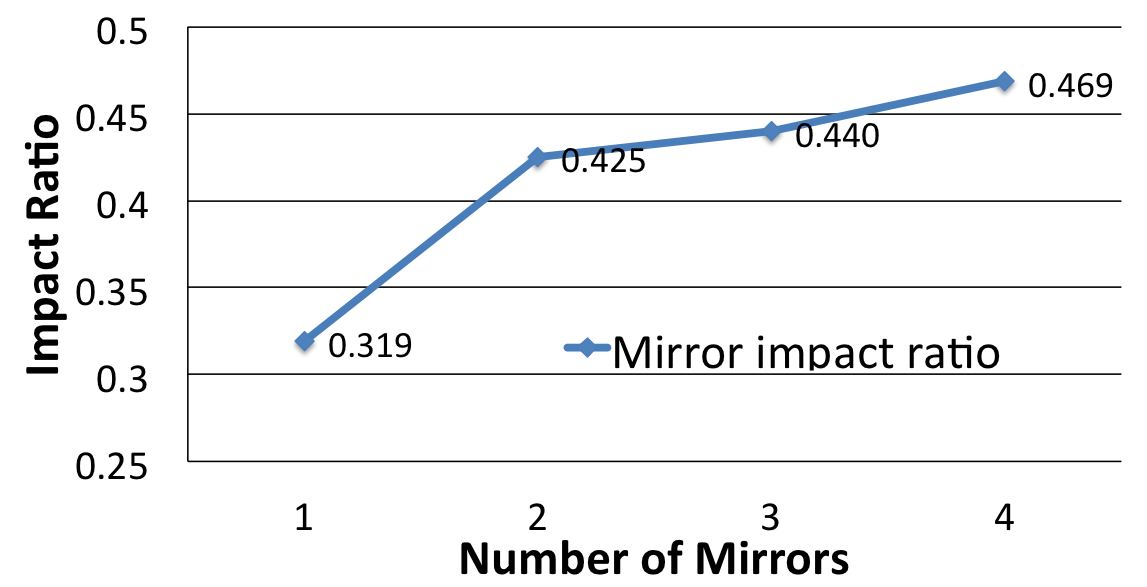} \end{center}
\vspace{-10pt}
\caption{Average impact ratio of the availability of users based on their number of mirrors over the entire duration of the experiment. \label{fig:impact_ratio_number} }
\end{figure}

We further consider the proportion of sessions that were handled by the mirrors instead of the user devices. We compute the {\em impact ratio}, which is the ratio of the number of sessions handled by mirrors to the total number of sessions for each group of users based on the number of their mirrors. As shown in Figure \ref{fig:impact_ratio_number}, having one mirror has around $32\%$ impact on profile availability while having $2$ mirrors increases this number to $42\%$ which means that a second mirror results in around $10\%$ increase in the overall profile availability. As for more than two mirrors the added availability is very small so we can conclude that there is an optimal number of mirrors based on the profile availability gained and in our experiments it is $2$. This number is a function of the overlap in terms of connectivity times of mirror devices and the mirrored user as well as the duration of their connectivity. In other words selecting mirrors that are mostly online during times that the original user is offline would decrease the number of mirrors one requires to have most profile availability. 
\comment{ 
As a future work we intend to integrate a recommendation system for MyZone client application that would analyze the connectivity times of friends and based on that, recommends the most appropriate friends as potential mirrors.
} 

\comment{ 
\begin{figure} \begin{center} \includegraphics[width=0.6\textwidth]{s2f5.png} \end{center}
\caption{Average impact ratio of the availability of users based on the rank of the mirror over the entire duration of the experiment. \label{fig:impact_ratio_rank} }
\end{figure}
} 

\comment{ 
As described in Section \ref{sec:applicationLayer}, when a user's primary device is offline all the requests for that user are going to be handled by its replicas starting at its secondary device and moving onto tertiary device and then first ranked mirror and so on until an available mirror is found. Figure \ref{fig:impact_ratio_rank} shows the impact of the mirrors based on their ranks on the profile availability of users. The impact ratio is computed in the same fashion as in Figure \ref{fig:impact_ratio_number} except that in here the impact ratio is computed for the rank of each mirror individually. As we can see, the first ranked mirrors are responsible for approximately $27\%$ of the gain in profile availability while second ranked mirrors are responsible for around $10\%$ and third and fourth ranked mirrors have very little impact. These results confirm with our earlier results that suggested an optimal number of mirrors for maximum gained availability. 
}

The important outcome from Figure \ref{fig:impact_ratio_number} 
%and \ref{fig:impact_ratio_rank} 
is that choosing the right mirrors is more important than the number of mirrors. This was illustrated by the fact that the added advantage of having more than $2$ mirrors in our experiment is very little. 

\begin{figure} \begin{center} \includegraphics[width=0.6\textwidth]{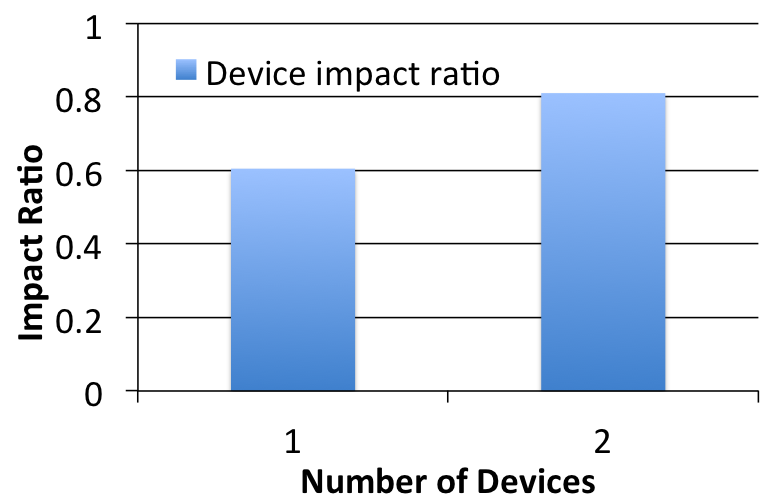} \end{center}
\vspace{-10pt}
\caption{The average impact ratio of the availability of users based on their number of devices over the entire duration of the experiment. \label{fig:impact_ratio_devices} }
\vspace{-5pt}
\end{figure}

Finally, we show the average impact of primary and secondary devices on profile availability in Figure \ref{fig:impact_ratio_devices}. As we can see, a second device would result in $20\%$ more impact on the availability, which is quite good
considering that a single device results in $60\%$ impact ratio. This shows that even if users can not obtain a mirror, having additional devices would have a significant impact on their availability. Of course when a user replicates her profile on several devices, these device connectivities are usually non overlapping. This means when one is connected the others are usually offline and overall this would increase the effectiveness of self replication. 

\subsection{Resource Utilization}
\label{sec:resource_utilization}

In MyZone, if a user chooses to be a mirror for another friend, her resource consumption will increase. This increase will generally be related to the number of friends and profile size of the mirrored friend. We analyze the storage and bandwidth requirements of MyZone client application related to mirroring. The bandwidth requirement of MyZone for a user profile will be analyzed later as part of scalability.

\begin{figure} \begin{center}\includegraphics[width=\textwidth]{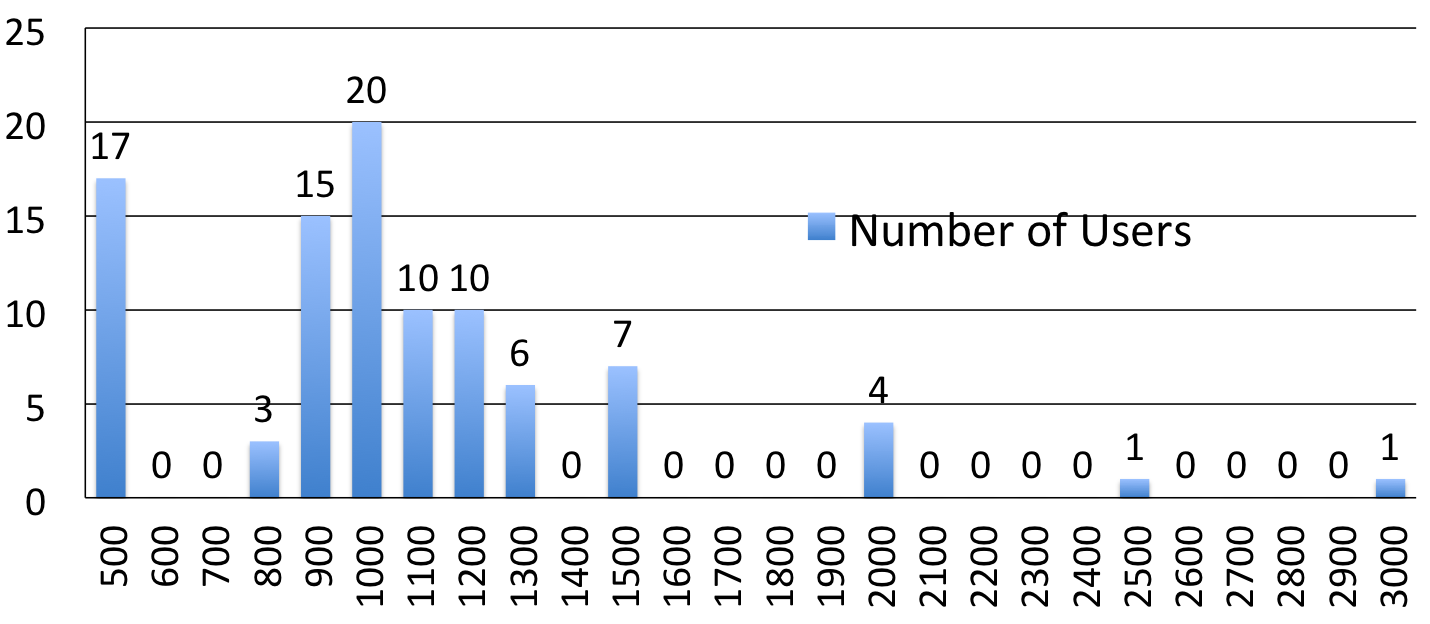} \end{center}
\vspace{-10pt}
\caption{Histogram of the number of users with respect to the mirroring storage capacity they are willing to allocate. \label{fig:allocated_storage} }
\vspace{-15pt}
\end{figure}

Figure \ref{fig:allocated_storage} shows the histogram of the number of users willing to be mirrors based on the storage that they are willing to allocate for mirroring. This figure shows that most users are willing to allocate $500$ MB to $1500$ MB of disk space to mirror their friends, which is quite low compared to the size of a typical hard drive. 
\comment{
For the very small number of users who are willing to allocate larger spaces to their mirrored friends, we have noticed that the social tie between the mirror and the mirrored users plays a considerable role in the willingness of the mirror.
}

\begin{figure} \begin{center} \includegraphics[width=\textwidth]{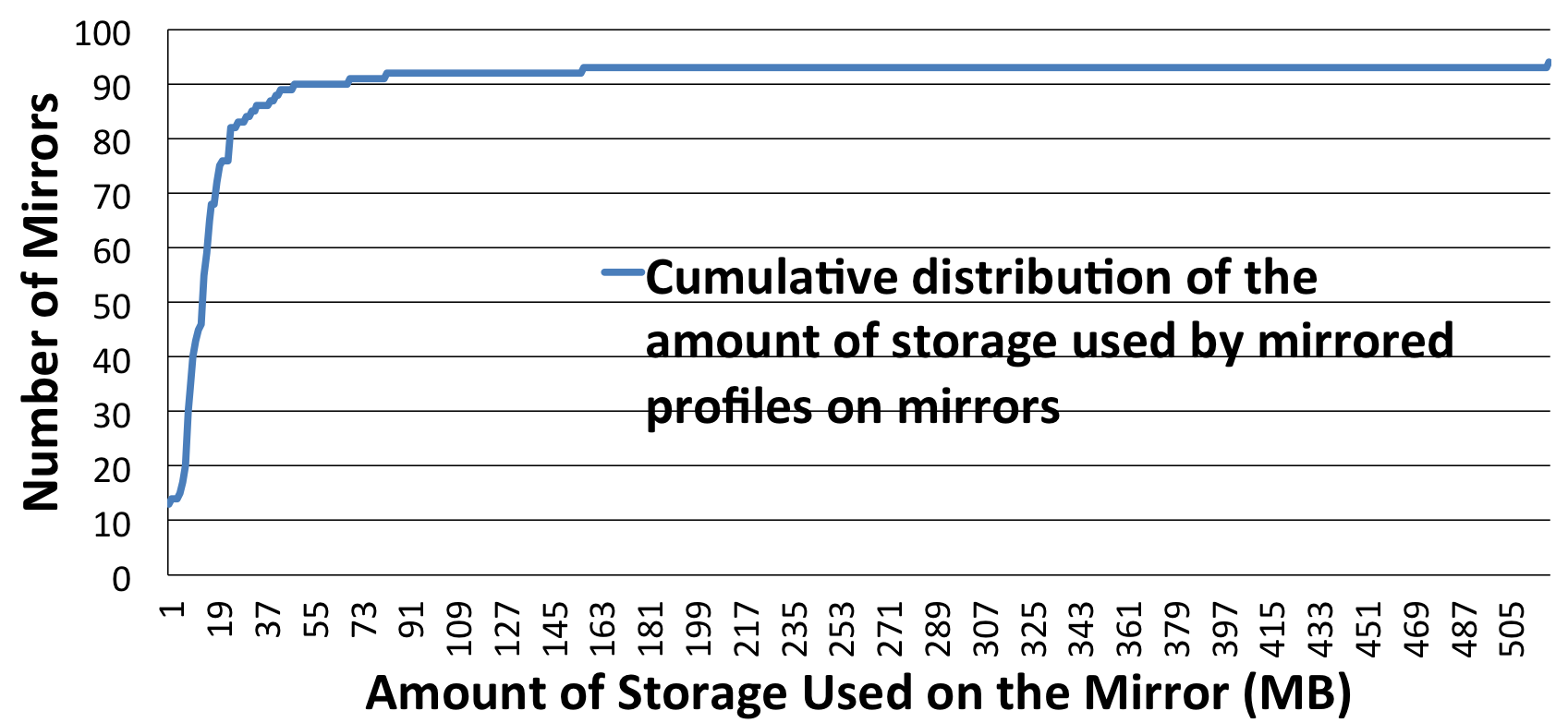} \end{center}
\vspace{-10pt}
\caption{Cumulative distribution of the number of mirroring users based on the actual size of the mirrored profiles. \label{fig:cdf_num_mirrors_profile_size} }
\vspace{-5pt}
\end{figure}

Next, we measure the actual size of the profiles stored on each mirror. Figure \ref{fig:cdf_num_mirrors_profile_size} shows the cumulative distribution of the number of mirrors based on the size of the mirrored profile. It indicates that most of the mirrored profiles occupy $3$MB to $23$MB of storage, and very few have much larger sizes. In fact $95\%$ of the mirrored profile sizes are less than $73$MB. This implies that the amount of disk space users are willing to allocate for mirroring is a significantly larger than the actual amount of storage needed for mirroring. Our results show that the tendency to allocate much larger stirage space for mirroring has strong correlation with self identified social ties among users. This indicates that users are not stingy when it comes to allocating space for mirroring people whom they have strong social ties with.
%which implies that users are willing to participate in social hosting. 

\begin{figure} \begin{center} \includegraphics[width=\textwidth]{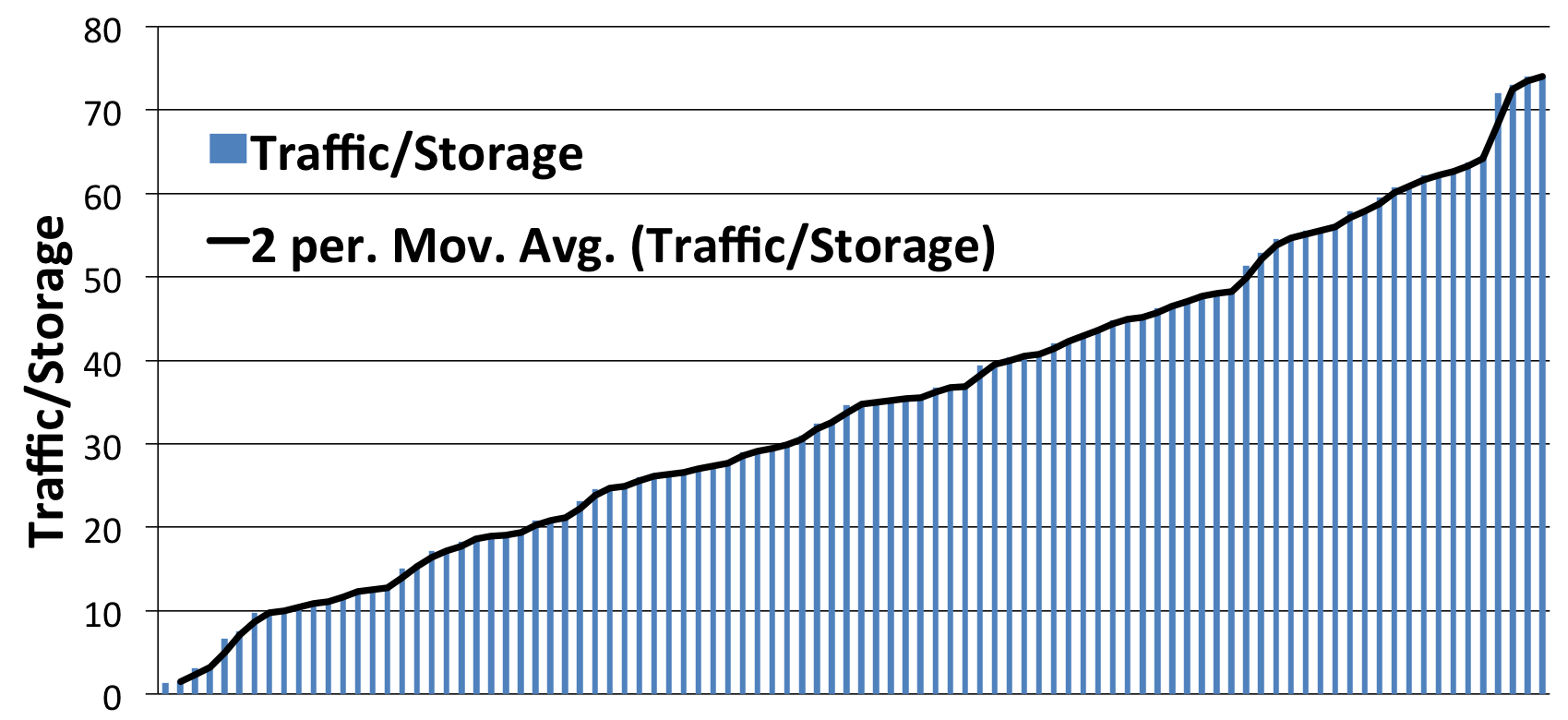} \end{center}
\vspace{-10pt}
\caption{The growth trend of traffic to used storage ratio for mirrored profiles. \label{fig:traffic_storage_growth} }
\vspace{-15pt}
\end{figure}

Next, we computed the ratio of the traffic generated by all mirrors of a profile to the size of the profile for all $94$ mirrored profiles. Figure \ref{fig:traffic_storage_growth} shows the ratios sorted in increasing order.
% This graph shows the growing trend of traffic to used storage for mirrored profiles.
The ratio of traffic to profile size for a mirrored user $x$ roughly indicates how many of $x$'s friends have used the mirror to interact with $x$. 
The graph indicates a linear behavior, where the ratios start at just over 1 and grow to just under 75. We investigated our log files and realized that this ratio is largely a function of the availability of the original user and the rank of the mirror (as will be discussed later) than anything else as almost all of our users had the same number of friends. Since the average number of friends in our social network is $98$, we can conclude from Figure \ref{fig:traffic_storage_growth} that a mirror is responsible for handling the requests of approximately $35$ friends of the mirrored user. This is because the total traffic to profile size should be equal to the number of friends for each user. This is due to the optimal traffic generation of MyZone as described in \ref{sec:applicationLayer}. Therefore, in total, mirrors are responsible for nearly $35\%$ of the total traffic in our experiment.

Finally, we have evaluated scalability aspects of MyZone based on cpu utilization and application response times. Due to space limitations we will not be able to present our results in details. However, our evaluation based on emulation of clients with up to $1600$ friends indicate that the MyZone client application can easily accommodate over a thousand friends per each user while the response time and the cpu utilization never exceed $500$ms and $70\%$ respectively.  The typical number of friends on popular OSNs is expected to be much lower~\cite{facebook_stat}.

%% file: contributions.tex
\section{Conclusion \& Future Work}
\label{ch:contributions}

\comment{The busy lifestyle of growing number of people has made conventional ways of socializing a luxury not available to all. Online social networks have made it possible for these people, to extend their social connections, while maintaining the existing ones. This has made OSNs a huge success with hundreds of millions of people using their services on daily basis. The large number of users has transformed OSNs into a more effective social media to spread ideas, news, opinions and more, comparing to conventional ones. 

Despite all their benefits, OSN service providers have been known to violate user privacy by selling user information to market researchers and other businesses. This has raised a lot of concerns in recent years, as people are becoming more and more aware of it and frequent changes in user privacy policies has only contributed to this. 

Furthermore, their growing use as a powerful social media has made them an effective tool to organize popular movements. This has made them a target of different attacks by the opposing entities. These attacks vary from traffic filtering to denial of service attacks to hijacking user information. 

Finally, ever-changing user interfaces and growing number of service providers have only made the user experience more frustrating. These shortcomings have motivated us to propose a P2P design for a next-generation OSN called MyZone. }

To our knowledge, MyZone is the first OSN with the following properties.  It preserves user privacy by storing user information on their own devices and replicating them on a number of other devices belonging to trusted friends. It is secure based on a {\it ``need to know basis''} philosophy and all connections are encrypted.  It is resilient to benign and malicious attacks.  The private social network model can be set up conveniently and quickly to construct a private OSN for a limited number of users.  It is backward compatible and a wide variety of client applications with different user interfaces and features can coexist while users can choose the applications that fit their needs and preferences.   We proved the feasibility of such a system by implementing the service layer part to support a private social network and a sample client application that supports common features of a conventional OSN. We presented detailed experimental results that analyze availability and resource utilization metrics for a deployment with over 100 users.  \comment{Based on our experimental results, we show that our proposed design can be a potential candidate to replace conventional OSN with the goal of providing more user privacy while being secure and resilient to attacks. }

%\subsection{Future Work}

\comment{The availability of user profiles even when the users are offline, was identified as perhaps the most serious obstacle faced by any P2P OSN. Our design addresses this by selecting mirrors among friends.}  We plan to develop a mirror recommendation system that selects the most appropriate friends as mirrors. This is a twofold problem, namely how do we select friends that results in maximum availability, and how do we incentivize friends to accept being mirrors given limited resources?  Providing the minimal incentives while increasing the chances of acceptance is analogous to the problem of bidding in online auctions.  We also plan to extend MyZone to a large scale public deployment and have built a Web site for public release~\cite{MyZoneWeb}.  We intend to support more mobile platforms as well, e.g. iPhone.

%% file: paper.bbl
\begin{thebibliography}{10}

\bibitem{diaspora}
\url{http://joindiaspora.com/}.

\bibitem{appleseed}
\url{http://appleseedproject.org/}.

\bibitem{peerbook}
\url{http://blogs.cs.st-andrews.ac.uk/peerbook}.

\bibitem{MyZoneWeb}
\url{http://www.joinmyzone.com/}.

\bibitem{peerson}
{\sc Buchegger, S., Schišberg, D., Vu, L., and Datta, A.}
\newblock Peer{S}o{N}: {P}2{P} social networking Ð early experiences and
  insights.
\newblock In {\em In Proc. ACM Workshop on Social Network Systems\/} (2009).

\bibitem{safebook}
{\sc Cutillo, L., Molva, R., and Strufe, T.}
\newblock Safebook: A privacy-preserving online social network leveraging on
  real-life trust.
\newblock {\em Communications Magazine, IEEE 47}, 12 (2009), 94 --101.

\bibitem{facebook_call}
{\sc Kincaid, J.}
\newblock Senators call out facebook on Ôinstant personalizationÕ, other
  privacy issues.
\newblock techcrunch.com, Apr. 27 2010.

\bibitem{cnn_egypt_censor}
{\sc Kravets, D.}
\newblock Twitter blocked in egypt amid street protests.
\newblock CNN.com, Jan. 26 2011.

\bibitem{MyZoneTechReportARXIV}
{\sc Mahdian, A., Black, J., Han, R., and Mishra, S.}
\newblock Myzone: A next-generation online social network.
\newblock {\em CoRR abs/1110.5371\/} (2011).

\bibitem{epic_fail}
{\sc Paul, R.}
\newblock {EPIC} fail: Google faces {FTC} complaint over buzz privacy.
\newblock artechnica.com, Feb. 17 2010.

\bibitem{washingtonpost}
{\sc Richburg, K.~B.}
\newblock Nervous about unrest, chinese authorities block web site, search
  terms.
\newblock WashingtonPost.com, Feb. 25 2011.

\bibitem{cnn_government}
{\sc Rushkoff, D.}
\newblock Internet is easy prey for governments.
\newblock CNN.com, Feb 5. 2011.

\bibitem{vis-a-vis}
{\sc Shakimov, A., Lim, H., Caceres, R., Cox, L.~P., Li, K., Liu, D., and
  Varshavsky, A.}
\newblock Vis-a-vis: Privacy-preserving online social networking via virtual
  individual servers.
\newblock In {\em 3rd Int. Conf. Comm. Sys. Net. (COMSNETS)\/} (2011), pp.~1
  --10.

\bibitem{facebook_stat}
{\sc Ugander, J., Karrer, B., Backstrom, L., and Marlow, C.}
\newblock The anatomy of the facebook social graph.
\newblock {\em CoRR abs/1111.4503\/} (2011).

\bibitem{nytimes_china_censor}
{\sc Wong, E., and Barboza, D.}
\newblock Wary of egypt unrest, china censors web.
\newblock NewYorkTimes.com, Jan. 31 2011.

\bibitem{cuckoo}
{\sc Xu, T., Chen, Y., Zhao, J., and Fu, X.}
\newblock Cuckoo: towards decentralized, socio-aware online microblogging
  services and data measurements.
\newblock In {\em HotPlanet '10\/} (New York, NY, USA, 2010), ACM,
  pp.~4:1--4:6.

\end{thebibliography}
